\documentclass[journal]{IEEEtran}
\ifCLASSINFOpdf
\else
\fi
\usepackage[tight,footnotesize]{subfigure}
\usepackage[pdftex]{graphicx}
\usepackage{amsmath}
\usepackage{amssymb}
\DeclareMathOperator*{\argmax}{arg\,max}

\begin{document}
%
\title{An Overview of Emerging Technologies for High Efficiency 3D Video Coding}
%
%
%

\author{Qifei~Wang
\thanks{Qifei Wang was with the Department of Electrical Engineering and Computer Sciences, University of California, Berkeley, CA 94720, USA, qifei.wang@eecs.berkeley.edu}
}

%
%

\markboth{Journal of \LaTeX\ Class Files,~Vol.~14, No.~8, August~2015}%
{Shell \MakeLowercase{\textit{et al.}}: Bare Demo of IEEEtran.cls for IEEE Journals}
%



\maketitle

\begin{abstract}
3D video coding is one of the most popular research area in multimedia. This paper reviews the recent progress of the coding technologies for multiview video (MVV) and free view-point video (FVV) which is represented by MVV and depth maps. We first discuss the traditional multiview video coding (MVC) framework with different prediction structures. The rate-distortion performance and the view switching delay of the three main coding prediction structures are analyzed. We further introduce the joint coding technologies for MVV and depth maps and evaluate the rate-distortion performance of them. The scalable 3D video coding technologies are reviewed by the quality and view scalability, respectively. Finally, we summarize the bit allocation work of 3D video coding. This paper also points out some future research problems in high efficiency 3D video coding such as the view switching latency optimization in coding structure and bit allocation.
\end{abstract}


%
\IEEEpeerreviewmaketitle

\section{Introduction}
%
%
%
%
\IEEEPARstart{H}{igh} immersive 3D visual content representation has been has been studied over decades. Nowadays, the 3D visual representation is applied in both professional fields (tele-immersive medicine and communications) \cite{wang2015unsupervised} and the personal applications (virtual reality and 3D computer gaming) \cite{zhang2007multiview}. In different applications, the 3D visual content has various representations, such as point clouds, meshes, multiview video (MVV), and texture video plus depth maps \cite{yemez2007scene}. For all these representations, the data used to represent the 3D visual content is significant more than the data of the 2D visual representation such as images and videos. Recent years, the worldwide research efforts have been spent on the data compression to improve the accessibility of the 3D visual content. For the video based applications, e.g. the MVV and free view-point video (FVV) \cite{smolic20113d}, the multiview video coding (MVC) \cite{merkle2007efficient} and some advanced 3D video coding technologies have ever been proposed based on the traditional video coding framework. For the graphics applications, like computer gaming and light field where the point clouds and meshes are usually applied, the 3D mesh coding \cite{peng2005technologies} has been widely studied. Especially, the scalable 3D mesh coding \cite{cao20123d} have gain interests from both academia and industry. Due to the increasing business in the 3D industry, ISO/IEC JTC 1/SC 29/WG 11 (Moving Picture Experts Group—MPEG) and the ITU-T SG 16 Working Party 3, the two main international organizations in the world which have worked individually and jointly, have worked on the standardization of the 3D video coding \cite{vetro2011overview} and 3D mesh coding. 

This paper reviews the recent progress of the coding technologies of 3D video. The 3D video mentioned in this paper mainly includes the binocular video, MVV, FVV which is represented in texture video plus depth maps. We reviews the coding framework of MVC and the joint coding technologies of MVV and depth maps. We also reviews the scalable 3D video coding frameworks for quality and view scalability. At last, the bit allocation in 3D video coding have been summarized. This paper also point out some unsolved problems, such as the optimization in coding structure and bit allocation when considering view switching behavior model.

The rest of this paper are organized as follows: Section \ref{sec:3D_representation} briefly introduces some background knowledge of 3D video representations, depth maps generation, and the view synthesis; Section \ref{sec:mvc} presents the MVC framework and its extensions to the FVV coding; Section \ref{sec:joint_coding} analyzes the joint coding technology of MVV and depth maps; Section \ref{sec:scalable_coding} reviews the work of both quality and view scalable 3D video coding; Section \ref{sec:bit_allocation} describes the bit allocation efforts of the FVV coding; The conclusions are summarized in Section \ref{sec:conclusion}.  

\section{3D Video Representations}
\label{sec:3D_representation}
In this section, we briefly introduce the background information about the 3D video representations, depth map generation, and virtual view synthesis. 

\subsection{Multiview Video and Free View-point Video}
\begin{figure}[t] 
	\centering
	\subfigure[Binocular Camera]{\label{fig:binocular_camera}
		\includegraphics[width=0.23\textwidth]{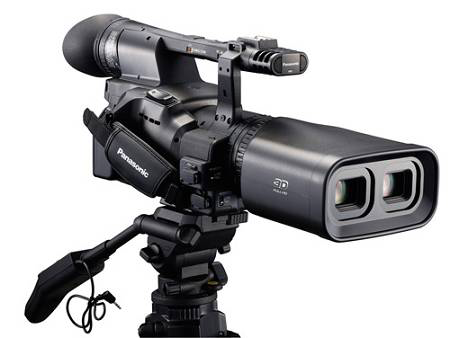}}
	\subfigure[MVV Camera]{\label{fig:mvv_camera}
		\includegraphics[width=0.23\textwidth]{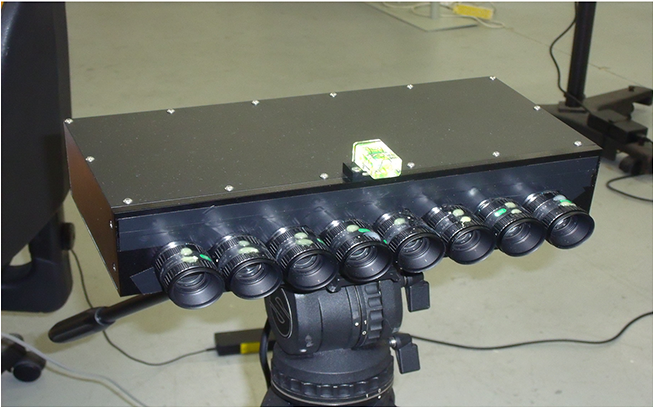}}
	\caption{Binocular and MVV camera systems.}
	\label{fig:camera_system}
\end{figure}

In the human vision system (HSV), the eyes capture the scene through their own channels. The brain merges the two streams of images and generates the stereo vision by the disparity between the images from different channels. Based on this observation, the stereoscopic video was proposed by presenting the videos from two parallel cameras to the two eyes respectively. Human can thus obtain stereo viewing experience when watching the stereoscopic video \cite{urey2011state}. The stereoscopic video can be captured by the binocular camera as shown in Fig. \ref{fig:binocular_camera}. Although the binocular video can generate stereo vision for HSV, the view point is still fixed. To enhance the immersive viewing experience, the MVV is proposed to add additional view points for the observer. The MVV are usually captured by the camera array systems as shown in Fig. \ref{fig:mvv_camera}. Compared to the binocular video, the MVV expands the vision field to the area captured by the camera array. Therefore, when watching MVV, the viewer can select any view point captured by the camera array. Moreover, the autostereoscopic video system demonstrates the video from multiple view points simultaneous on the polarized screen that viewer can watch the stereoscopic video from multiple physical positions. The MVV can though provide more immersive viewing experience than the binocular stereoscopic video, the view point is nevertheless limited to the captured ones. In order to obtain full immersive viewing experience, people proposed the FVV \cite{tanimoto2011free} which requires to provide arbitrary view point accessibility. However, it is physically hard to obtain dense sampling of the whole 3D space. Therefore, the FVV is usually achieved by the virtual view synthesis with the sparsely captured texture video and geometry information. The geometry information can be represented in multiple different formats, such as depth, point cloud, meshes, voxels, etc. \cite{yemez2007scene}. For the purpose of data streaming and real-time processing, the depth maps are the most popular format used in FVV. Therefore, the FVV is usually represented by the MVV which is captured by the sparse camera array and the depth maps \cite{muller20113}. 

\subsection{Depth Map Generation}
The depth maps which represent the quantized distance from the object to the camera plane in the 3D space can be generated by multiple ways. The highest precision depth maps are usually generated by the stereo matching between multiple images from different views \cite{sun2005symmetric}. Some recent research can also generates the depth maps from a single image by exploiting the implicit geometry prior within the texture images \cite{saxena2005learning}. Besides the computational approaches, the depth maps can also be sampled by the depth sensors. The depth sensors can be categorized into three classes by their sensing principles, including structure light \cite{wang2015computational} (Microsoft Kinect version 1), time-of-flight infrared camera \cite{zhang2012microsoft} (Microsoft Kinect version~2), and the laser scaning (MC3D) \cite{matsuda2015mc3d}. Although all these depth sensors are suffered from the noise and low resolution \cite{wang2015evaluation}, they can capture the depth maps in real-time. Therefore, for the real-time applications, such as the interactive computer gaming, the depth sensors are widely applied. One the other hand, for the applications like 3DTV which requires high precision depth information, the computational depth estimation approaches are usually applied.

\subsection{Virtual View Synthesis}
\begin{figure}[htbp] 
	\centering
	\includegraphics[width=0.4\textwidth]{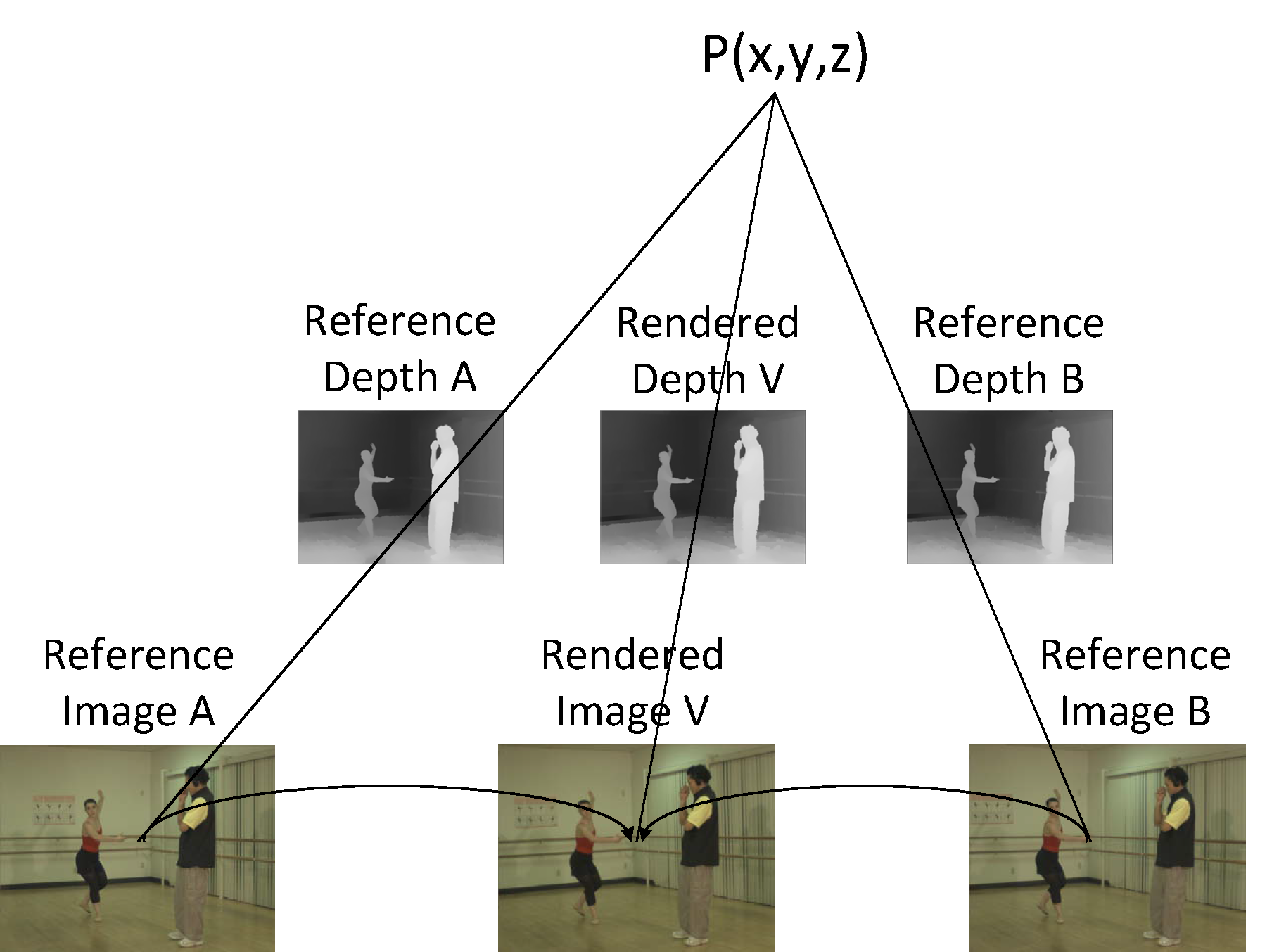}
	\caption{Virtual view synthesis.}
	\label{fig:rendering}
\end{figure}

In the early stage of the 3D video, the virtual view video is interpolated from the texture images from its reference views by the coarse geometry models. However, without the accurate epipolar geometry model, the interpolation usually cause artifacts on the objects with large disparities. In the latest view synthesis framework, based on the epipolar geometry model and the calibrated camera parameters, each pixels in the reference view can be projected to the virtual view plane with its depth value \cite{chan2007image}. In the FVV, the virtual view synthesis is implemented based on the latest view synthesis framework as shown in Fig. \ref{fig:rendering} where the virtual view is synthesized by its left and right reference views. In order to reduce the pixel synthesis drift caused by the depth noise, the view synthesis module usually synthesize the depth maps of the virtual view with the depth maps from the reference views. Afterward, each pixel in the virtual view is backward mapped to the reference views and interpolated with the neighbors if the corresponding position is out of the sampled grid in the reference views. The backward projection reduces the ambiguity in the pixel merge from multiple reference views.

\section{Simulcast and Multiview Video Coding}
\label{sec:mvc}
\begin{figure*}[t]
	\centering     
	\subfigure[Simulcast]{\label{fig:mvc_simulcast}
		\includegraphics[width=0.3\textwidth]{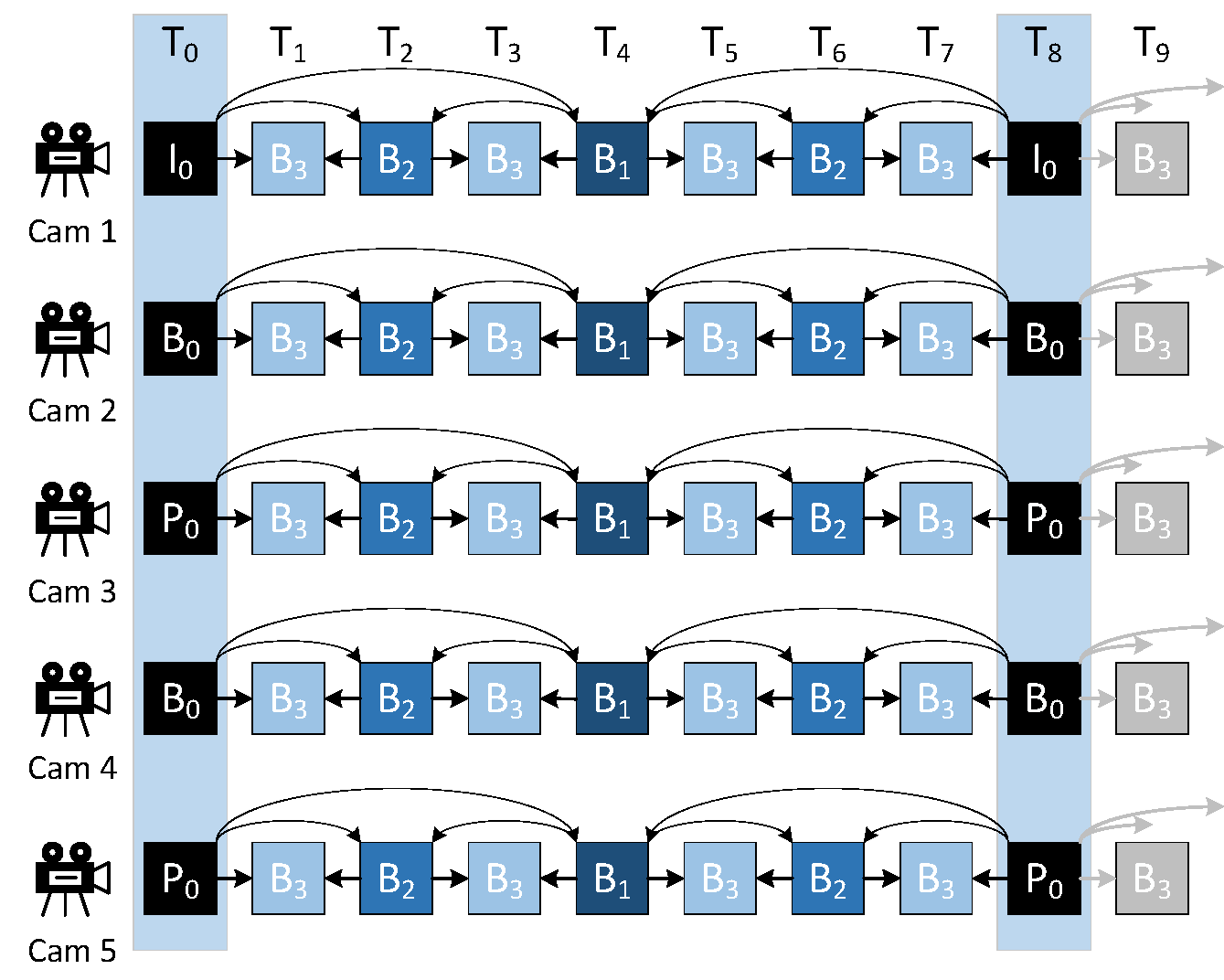}}
	\subfigure[MVC\_KS]{\label{fig:mvc_ks}
		\includegraphics[width=0.3\textwidth]{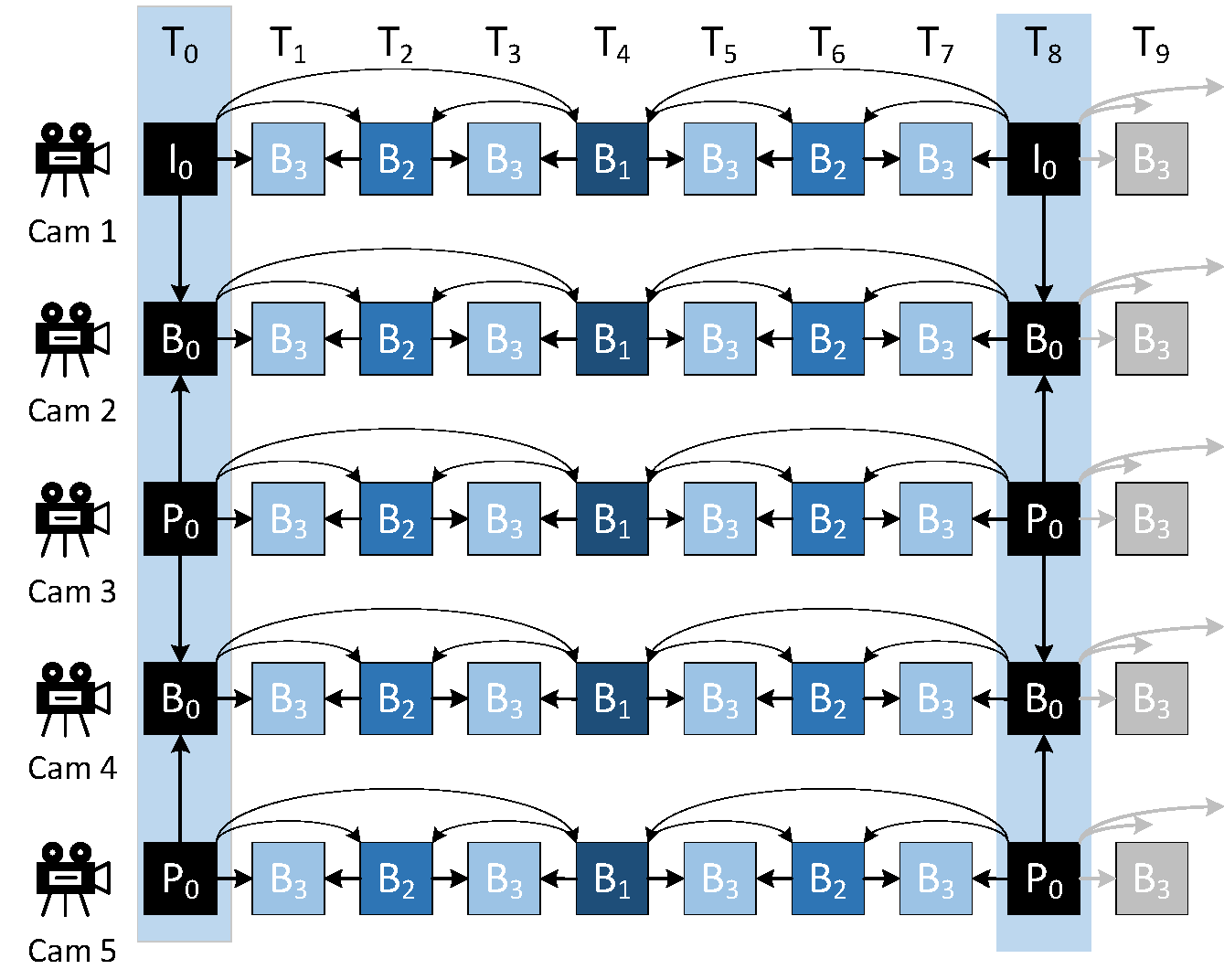}}
	\subfigure[MVC\_AS]{\label{fig:mvc_as}
		\includegraphics[width=0.3\textwidth]{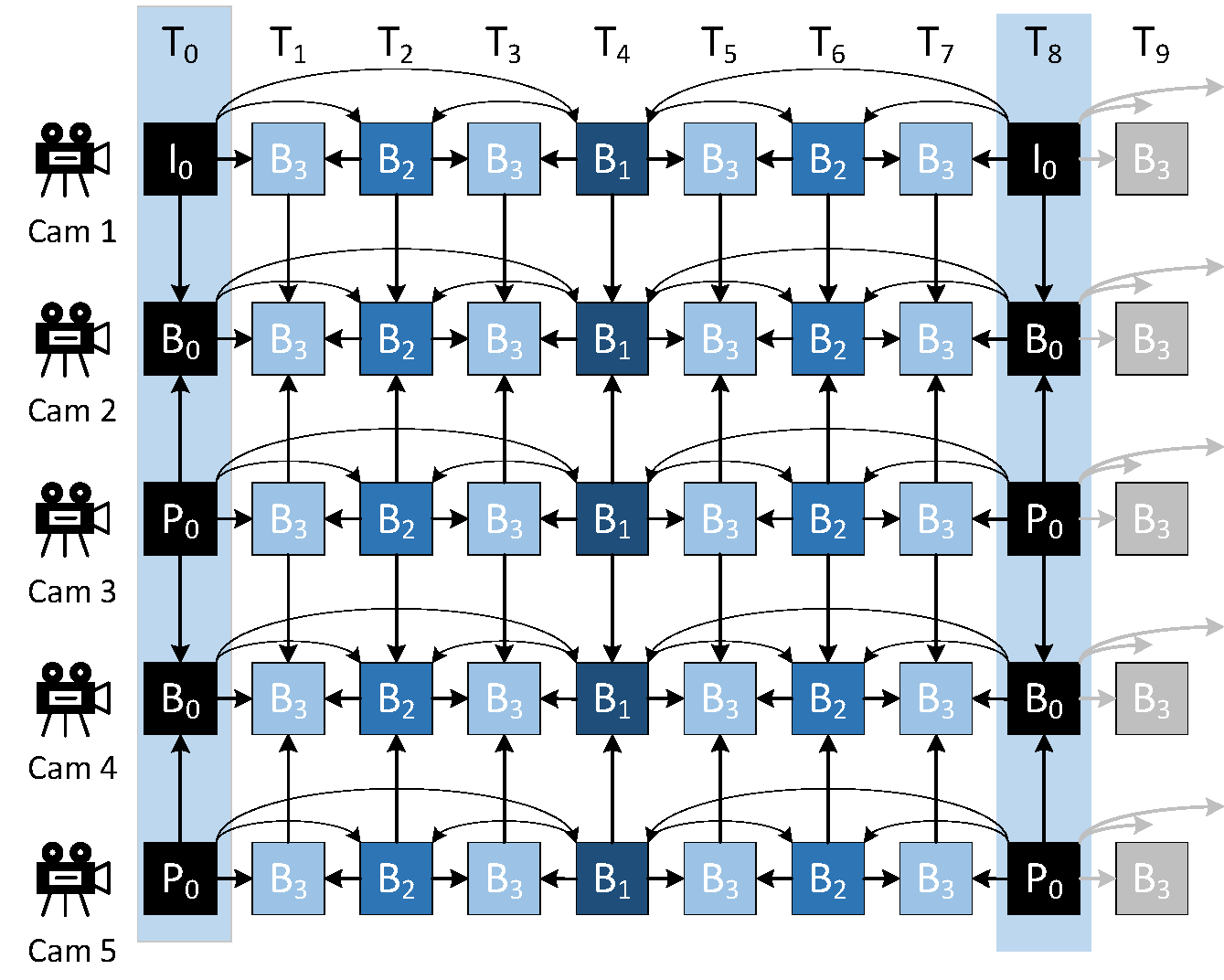}}
	\caption{Prediction structure in the MVC framework.}
	\label{fig:mvc}
\end{figure*}

As the MVV is formed by the video stream from multiple view points, each video stream can be encoded individually by single view video codec. This encoding framework is called simulcast as shown in Fig. \ref{fig:mvc_simulcast}. Although the single view video codec can exploit the temporal and intra frame correlation, the redundancy between different views still remains in the simulcast video streams. In order to compress the redundancy between different views, the multiview video coding (MVC) framework is proposed by adding the inter-view prediction to the simulcast coding framework. Fig. \ref{fig:mvc_ks} demonstrates the MVV coding framework with the inter-view prediction implemented on the key frames which is called MVC\_KS in the rest of this paragraph. The coding framework in Fig. \ref{fig:mvc_as} extends the inter-view prediction to both key and non-key frames to fully exploit the inter-view redundancy. The coding framework in Fig. \ref{fig:mvc_as} is called MVC\_AS in the rest of this paragraph. 

\begin{figure}[t] 
	\subfigure[Dancer]{\label{fig:mvc_rd_dancer}
		\includegraphics[width=0.23\textwidth]{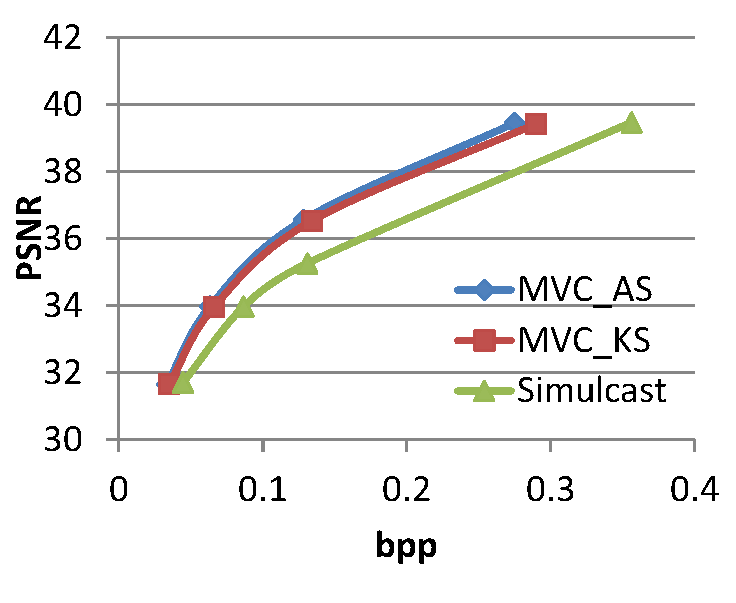}}
	\subfigure[Poznan\_Street]{\label{fig:mvc_rd_pstreet}
		\includegraphics[width=0.23\textwidth]{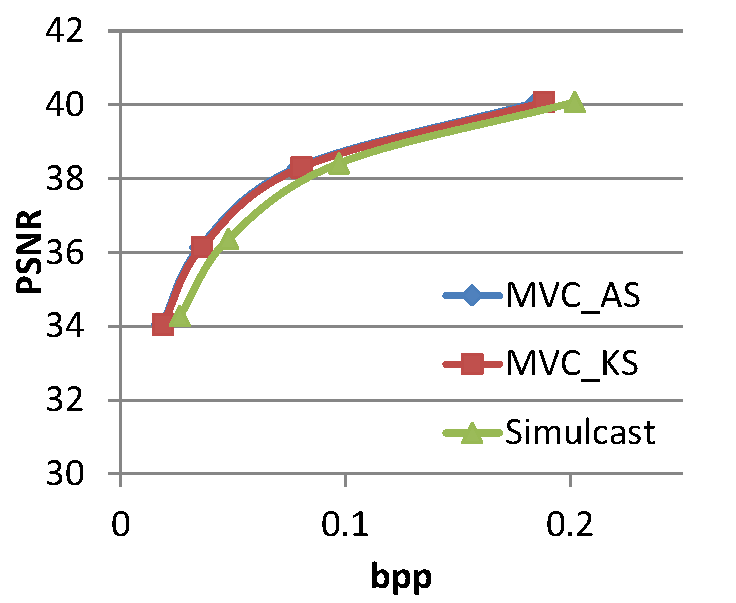}}
	\caption{R-D performance of simulcast, MVC\_KS, and MVC\_AS.}
	\label{fig:mvc_rd}
\end{figure}

Fig. \ref{fig:mvc_rd} demonstrates the rate-distortion (R-D) performance of different coding frameworks. Two testing sequences, "Dancer" and "Poznan\_Street" which are both at 1080p and 30 fps, are used in this paper for all the evaluation experiments. From the curves shown in Fig. \ref{fig:mvc_rd}, we can see the MVC frameworks improve the coding efficiency compared to the simulcast coding framework. The coding structure with full inter-view prediction (Fig. \ref{fig:mvc_as}) achieves further R-D improvement than the coding structure with inter-view prediction on only key frames. 

For the purpose of compressing MVV, the MVC\_AS framework would be the optimal choice. However, in the FVV applications, the viewer may randomly switch the view point. Although the simulcast coding framework achieves the lowest R-D performance compared to the other two, the video stream of different view point can be randomly accessed without any dependency. The MVC\_KS framework achieves both moderate R-D performance and the view switching latency. The MVC\_AS framework achieves the best coding framework at the cost of the highest view switching latency. Therefore, the prediction structure should be optimized to balancing the R-D performance and the view switching latency \cite{ji2014online}.

Based on the coding structure shown in Fig. \ref{fig:mvc}, the average view switching latency can be modeled by the following model. Denote there are $N$ frames in each group of pictures (GOP). The deepest level of the interview dependency is $M$. Suppose the view switching is fully random and satisfied the uniform distribution among all the frames and views, the average view switching delay $T$ can be represented by 

\begin{equation}
	T(M, N)=c\times M \times N.
	\label{equ:delay}
\end{equation}
In equation \ref{equ:delay}, $c$ denotes a constant coefficient which means the average view switching delay is proportional to the product of the GOP size and the average depth of the inter-view dependency. Consequently, the overall optimization problem can be represented as

\begin{equation}
	\begin{split}
	J(M,N,Q) &=\argmax_{M,N,Q}D(Q)+\lambda R(M,N,Q)+\beta T(M,N) \\
	R(M,N,Q) &< R_C
	\end{split}	
	\label{equ:optimization}
\end{equation}
In equation \ref{equ:optimization}, $D$ and $R$ denote the coding distortion and bit rate, respectively, $Q$ denotes the quantization parameter, $\lambda$ and $\beta$ denote the Lagrange multipliers, $R_C$ denotes the given bit rate constraint. Increasing $M$ and $N$ can reduce the coding bit rate but also introduce additional the view switching latency. On the other hand, when $M$ and $N$ are small, the system can obtain good random view accessibility with the cost of increasing bit-rate. This optimization problem can be solved by the Karush–Kuhn–Tucker (KKT) conditions.
 
\section{Multiview Video and Depth Maps Coding}
\label{sec:joint_coding}
After introducing the depth maps into the 3D video for the virtual view rendering, the depth maps coding technologies are further considered by the 3D video coding communities. Since the depth information are usually represented by the gray-level pictures, the depth maps are treated as the monochromatic MVV sequences that can be encoded by the MVC technologies. Therefore, most of 3D video coding frameworks encode the MVV and depth maps separately by MVC. Besides, the other existing depth maps coding technologies encode the depth maps by considering the signal properties of the depth maps \cite{oh2011depth}.

Although the MVC can remove both the intra-view and inter-view redundancy in both the MVV and depth maps, the correlation between the MVV and depth maps cannot be exploited by the MVC framework. Therefore, various joint multiview video and depth map coding technologies have been proposed to improve the 3D video coding efficiency based on MVC framework. Since the depth information can be converted to the disparity, using the depth maps to assist the inter-view prediction is one direction to improve the coding efficiency. In \cite{yea2009view}, Yea et al. proposed an auxiliary view synthesis prediction (VSP) mode to improve the inter-view prediction efficiency. In the VSP mode, an additional reference frame is generated for each coding frame by warping the pictures in the reference view to the current view with the decoded depth maps. The rendered virtual reference frame is added into the reference frame buffer for the prediction. Compared to the reference frame from the reference view, the rendered virtual reference frame is in the same camera plane as the current frame. Therefore, the estimated disparity vector between the blocks in the current frame and the virtual reference frame is much less than that between the current frame and the reference frame from reference view. Therefore, the VSP mode can reduce the R-D cost on the disparity vector compared to the traditional inter-view prediction mode. On the other hand, if the non-translational transform exists between different views, the VSP mode can warp the reference frame to the same camera plane as the current frame. This can significantly reduce the prediction residue compared to the translational disparity compensation prediction (DCP). 

\begin{figure}[t] 
	\subfigure[Dancer]{\label{fig:dadcp_dancer}
		\includegraphics[width=0.23\textwidth]{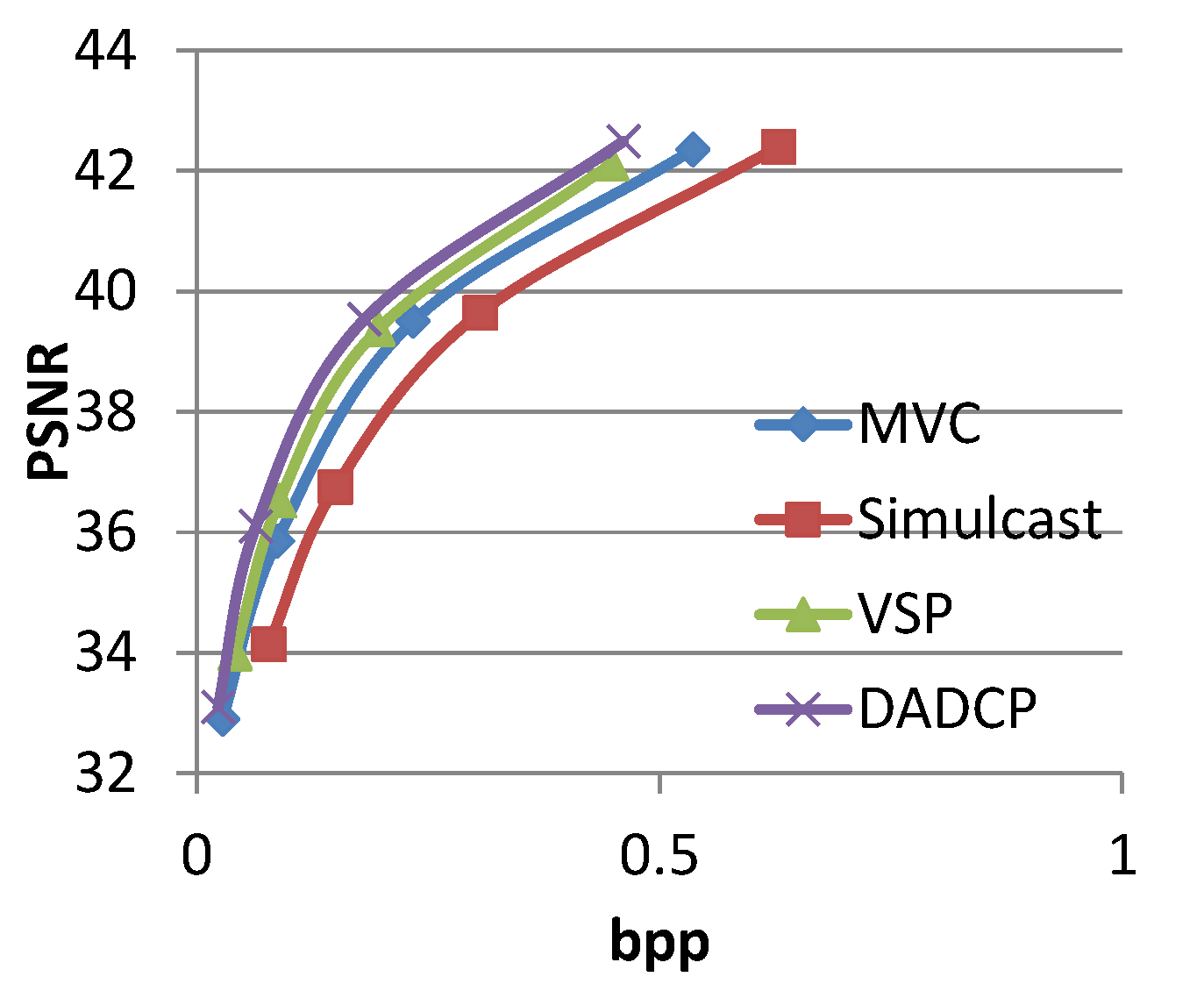}}
	\subfigure[Poznan\_Street]{\label{fig:dadcp_pstreet}
		\includegraphics[width=0.23\textwidth]{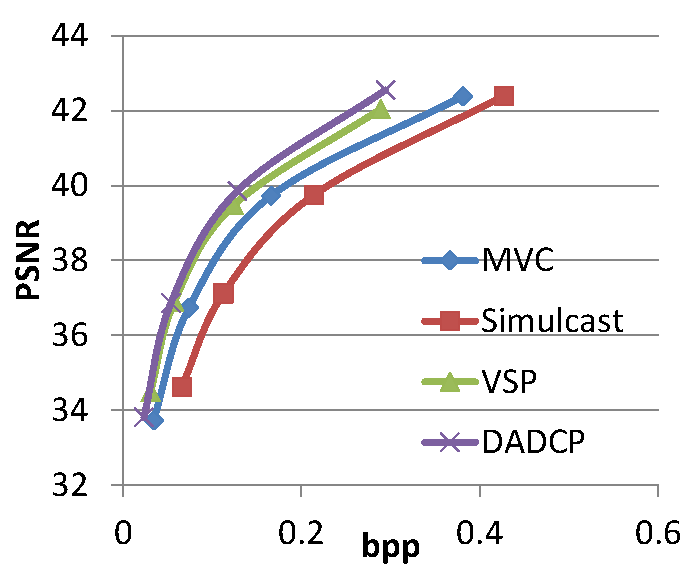}}
	\caption{R-D performance of the joint MVV and depth maps coding techniques.}
	\label{fig:joint_coding_rd}
\end{figure}

Although the VSP mode can reduce the bit-rate in inter-view prediction, the additional buffer cost increase the complexity at both encoder and decoder sides. In order to exploit the depth information to assist inter-view prediction, Wang et al.\cite{wang2012free} proposed a depth-assisted disparity compensation prediction (DADCP) mode. In this mode, the encoder calculates the disparity vector for each pixel in the block based on the calibrated camera parameters and the depth value of each pixel. During the prediction, the disparity vectors calculated from the depth maps are quantized to quarter pixel precision. The prediction residue of the DADCP mode is generated by per-pixel interview prediction. Since the depth maps can be treated as the side information for the texture video coding if the depth maps are encoded/decoded ahead of the multiview video encoding/decoding, the bit-rate of the depth calculated disparity vectors can thus be saved. The DADCP scheme can also save the bit-rate of disparity vector without the additional buffer cost compared to the VSP. Besides, since DADCP does not require any additional motion search in the prediction, its run-time complexity is also lower than that of the VSP mode. Figs. \ref{fig:dadcp_dancer} and \ref{fig:dadcp_pstreet} demonstrate the R-D performance of the MVC, MVC with VSP, and MVC with DADCP on the two testing sequences, respectively. From the curves, we can see that DADCP outperforms both the traditional MVC and MVC with VSP. The R-D gain is increasing with the bit-rate rising. Compared to the traditional MVC, the gain of DADCP comes from the bit rate reduction on the disparity vectors in the inter-view prediction. Moreover, the rendering distortion of the synthesized virtual reference frame suppresses the R-D performance of the VSP mode.

Beside using depth to assist the inter-view prediction, other research works exploit the similarity of the motion vector field between the MVV and depth maps to reduce the overall bit-rate of 3D video coding. Despite the lack of the texture, the depth maps record the same scene as the texture video. The depth maps and texture video can share similar motion vector fields. Zhang et al. \cite{zhang2010joint} proposed a joint coding framework that uses the motion vectors of the texture video as a candidate reference motion vector for the depth maps coding. However, since the depth maps are usually noisy than the texture video, the optimize motion vectors of the depth maps usually differ from those the corresponding block in the texture video. The R-D gain of the motion vectors sharing between texture video and depth maps is limited. Guo et al. \cite{guo2006inter} proposed an inter-view direct mode based on the parallelogram constraint between the motion and disparity vectors. The parallelogram constraint comes from the conservation of the temporal and spatial optical flow which means in any two pictures from one view and the other two corresponding frames from another view, the two motion vector and two disparity vectors between the corresponding pixels form a parallelogram. This scheme can reduce the coding bit-rate of the reference motion or disparity vector also save the run-time complexity at the encoder side.

Due to the smoothness of the depth maps, some efforts have been spent on down-sampling the depth maps to lower resolution and perform up-sampling after the decoding \cite{oh2009depth}. Since the bit-rate of the depth maps is usually 10\% to 20\% of the bit-rate of the texture video, the bit-rate savings on depth maps are limited compared to that of the texture video. Majority of the work still focuses on reducing the bit-rate of the MVV by the joint coding techniques. 

Due to the advantages of geometry block partitioning, Wang et al. \cite{wang2011reduced} proposed practical geometry block partitioning scheme by exploiting the correlation of the object boundary between texture and depth images. The proposed scheme searches the partition line by a linear operator based on the input texture and depth information which significantly reduce the complexity of the geometry partitioning. By enabling the geometry partitioning, the proposed coding framework can achieve 6\% R-D gain compared to the traditional MVC framework with only 18\% encoding time increasing \cite{wang2013complexity}. The proposed partition linear searching scheme can also be extended to the traditional 2D video compression \cite{wang2012complexity}. 

For the binocular stereoscopic video coding, some researchers have proposed a frame compatible 3D video coding framework by merging the frames from different views into a single frame. The interlacing mode includes top-bottom, side-by-side, row-interleaved, column-interleaved, and checkerboard \cite{vetro2010frame}. The sequences of the interleaved images can be compressed by the traditional single view video codec.

\section{Scalable 3D Video Coding}
\label{sec:scalable_coding}
\begin{figure}[t] 
	\centering
	\subfigure[Dancer]{\label{fig:svc_dancer}
		\includegraphics[width=0.23\textwidth]{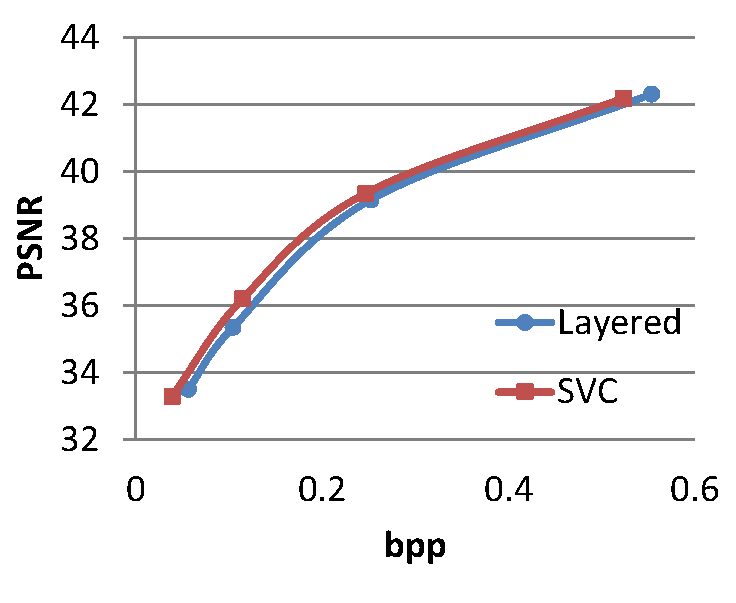}}
	\subfigure[Poznan\_Street]{\label{fig:svc_pstreet}
		\includegraphics[width=0.23\textwidth]{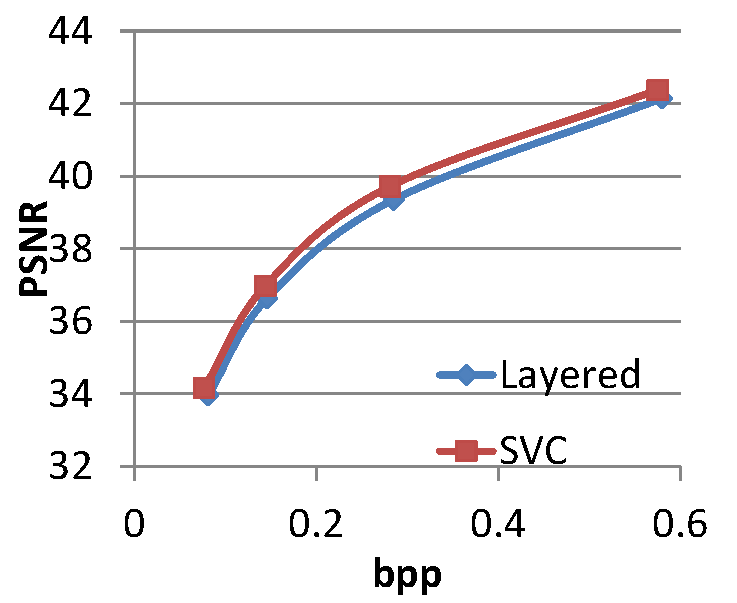}}
	\caption{R-D performance of the scalable 3D video coding techniques.}
	\label{fig:svc_rd}
\end{figure}

Due to the limited bandwidth and the high coding bit-rate of 3D video, the scalable 3D video coding also gains much research attention recently. In \cite{kurutepe2007client}, Kurutepe et al. proposed a resolution scalable MVC framework. In this framework, the encoder down-samples all the texture frames to a lower resolution. The base layer stream is generated by encoding all the texture frames at low resolution by MVC. The enhanced layer is the residue between the original frames and the decoded base layer frames. For each view, the enhanced layer is encoded independently by the traditional single view video codec. The encoder will deliver all the base layer and the enhanced layer of the views selected by the viewer. If the viewer change the view points, the enhanced layer can be switched with very low latency since there is no inter-view dependency in the enhanced layer. Therefore, this framework can also be applied to the scalable coding of the FVV. This framework is further extended to the quality scalable MVC framework in \cite{ji2014online} by applying the traditional scalable coding technology in MVC. Figs. \ref{fig:svc_dancer} and \ref{fig:svc_pstreet} demonstrates the R-D performance of these two coding frameworks on the two testing sequences, respectively. From Fig. \ref{fig:svc_rd}, we can see that the quality scalable framework outperforms the resolution framework since former framework applies the interlayer prediction when encoding the enhanced layer.

In MVV and FVV, the view scalability is a new research direction in 3D video coding. Shimuzu et al. \cite{shimizu2007view} proposed a view scalable MVC framework based on the view synthesis. In their framework, the video sequence of one view is selected as the base view and encoded by the traditional single view codec. The rest views are all treated as the enhanced views. The frames in the enhanced views are predicted by the corresponding frames in the base view. In this framework, the decoder obtains fully scalability of the view switching since all the views can be decoded based on the base view and its own prediction residue. However, the coding efficiency is much lower than the MVC framework. To further improve the coding efficiency, the temporal prediction can also be applied to the enhanced views. The random view accessibility is still maintained since all the views only depend on the base view. In this framework, the number of the base views and the prediction between the base views and the enhanced views can be extended. For example, the hierarchical inter-view prediction structure can also be applied in this framework to improve the view scalability. In the 3D video coding system, the coding framework can be optimized by balancing the coding efficiency and the view scalability.

\section{Bit Allocation of 3D Video Coding}
\label{sec:bit_allocation}
In FVV, since the video sequences displayed at the client side are synthesized by the texture video sequences and the depth maps of the reference view points, the quality of the synthesized video are determined by both the quality of the decoded texture video and depth maps. Based on the principle of virtual view synthesis, the distortion of texture video will introduce linear error to the synthesized video frames. On the other hand, the distortion of depth maps will results in the pixel position drifting error to the synthesized video frames. To obtain the optimized quality fo the synthesized video, the encoder has to optimize the bit allocation between the video and depth maps. In \cite{liu2009joint}, Liu et al. proposed a frame-level view synthesis distortion estimation algorithm and designed the bit allocation algorithm based on the rate distortion model of 3D video coding. However, since the frequency domain view synthesis distortion model needs the region has the uniform disparity error, the frame-level model is inaccurate to generate the R-D model. To get an accurate R-D model of 3D video coding, Wang et al. \cite{wang2010region} proposed a region based view synthesis distortion estimation model by partitioning the whole frame into the region with uniform depth value. The distortion estimated for each depth-uniform region can significantly reduce the error of the view synthesis distortion estimation. Based on the proposed R-D model, Wang et al. \cite{wang2012free} proposed a bit allocation algorithm with single pass search. The results reported in \cite{wang2012free} demonstrated that the region based R-D model can obtain more accurate estimation of the view synthesis distortion and improved R-D performance of the synthesized virtual view video compared to the frame-level R-D model. Besides, Yuan et al \cite{yuan2011model}. solved the bit allocation problem by the Lagrange optimization which can optimize the R-D performance with very low complexity.

Besides the bit allocation between the MVV and depth maps, some other works have studied the bit allocation between different views. In \cite{shao2012asymmetric}, Shao et al. proposed an asymmetric stereoscopic video coding scheme and optimize the bit allocation based on the masking effect of HSV. Yuan et al. \cite{yuan2015rate} proposed the bit allocation algorithm between different view based on the view switching. However, since the view switching behavior is complicated, considering the view switching behavior model in the bit allocation is still an unsolved problem for the free view-point video coding with multiple input view points.

\section{Conclusions}
\label{sec:conclusion}
In this paper, we reviewed the recent progress of the high efficiency 3D video coding technologies. Most of the MVV and FVV coding are based on the MVC framework. For the FVV represented by MVV and depth maps, majority research work focus on the joint MVV and depth maps coding techniques to reduce the overall coding bit rate. By considering the view switching, the prediction structure of MVV and FVV needs to be optimized to balance the view switching latency and the R-D performance. To make the 3D video streaming adaptive to the bandwidth and viewing behavior model, the scalable MVC and FVV coding are also an important research area for exploring. At the end, the bit allocation and rate control also need to be further optimized by considering the viewing behavior model. With the emerging applications of 3D video, the coding technologies will be further improved to meet the requirements from various applications.

\bibliographystyle{IEEEtran}
\bibliography{3DVCoding}

%


\begin{IEEEbiographynophoto}{Qifei Wang}
received the B.S. degree in information and computing science from Beijing University of Posts and Telecommunications, China, in 2007, and Ph.D degree in control science and engineering from Tsinghua University, China, in 2013. He joined EECS, University of California, Berkeley, US, in 2014. His current research interests include computer vision, machine learning, video processing and communications.
\end{IEEEbiographynophoto}






\end{document}